\documentclass[12pt]{iopart}

\usepackage{graphicx}
\usepackage{epsfig,latexsym}

\newcommand{\be}{\begin{eqnarray}}
\newcommand{\ee}{\end{eqnarray}}

\begin{document}

\title[S-Wave Quarkonia in Potential Models]{S-Wave Quarkonia in Potential Models}

\author{\'Agnes M\'ocsy and P\'eter Petreczky}

\address{RIKEN-BNL Research Center \& Physics Department, \\ Brookhaven National Laboratory, Upton, NY 11973, USA}
%\ead{mocsy@bnl.gov}
\begin{abstract}
We discuss S-wave quarkonia correlators and spectral function using the Wong-potential, and show that these do not agree with the lattice results. 
\end{abstract}

%Uncomment for PACS numbers title message
%\pacs{00.00, 20.00, 42.10}
% Keywords required only for MST, PB, PMB, PM, JOA, JOB? 
%\vspace{2pc}
%\noindent{\it Keywords}: Article preparation, IOP journals
% Uncomment for Submitted to journal title message
%\submitto{\JPA}
% Comment out if separate title page not required
%\maketitle

For the last twenty years the melting of heavy quark bound states has been considered an unambiguous signal for deconfinement \cite{Matsui:1986dk}. Understanding the modification of the properties of heavy quarkonia in a hot medium is therefore essential for understanding deconfinement. Phenomenological studies based on potential models have been recently accompanied by lattice QCD calculations. 

\begin{figure}[htbp]
\hspace*{1.2cm}
\begin{minipage}[htbp]{6cm}
\epsfig{file=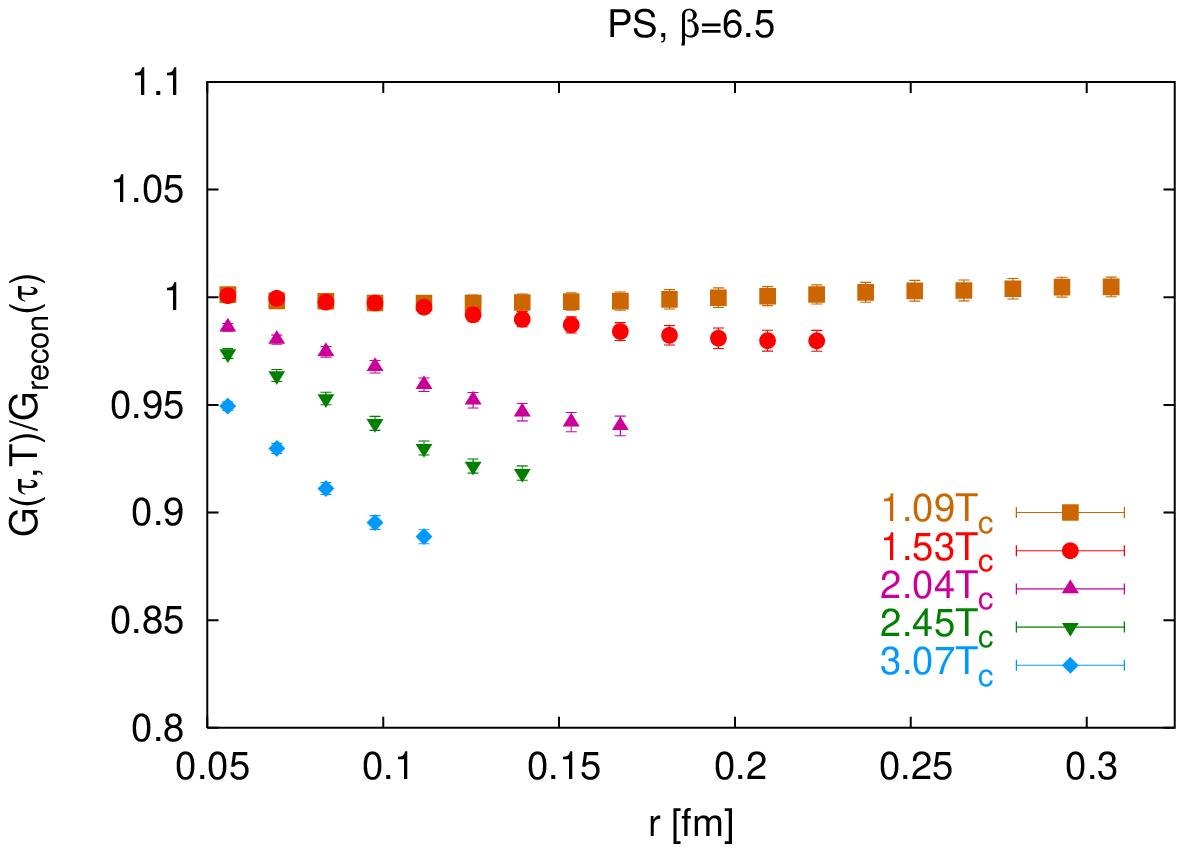,height=43mm}
\end{minipage}
\hspace*{1cm}
\begin{minipage}[htbp]{6cm}
\epsfig{file=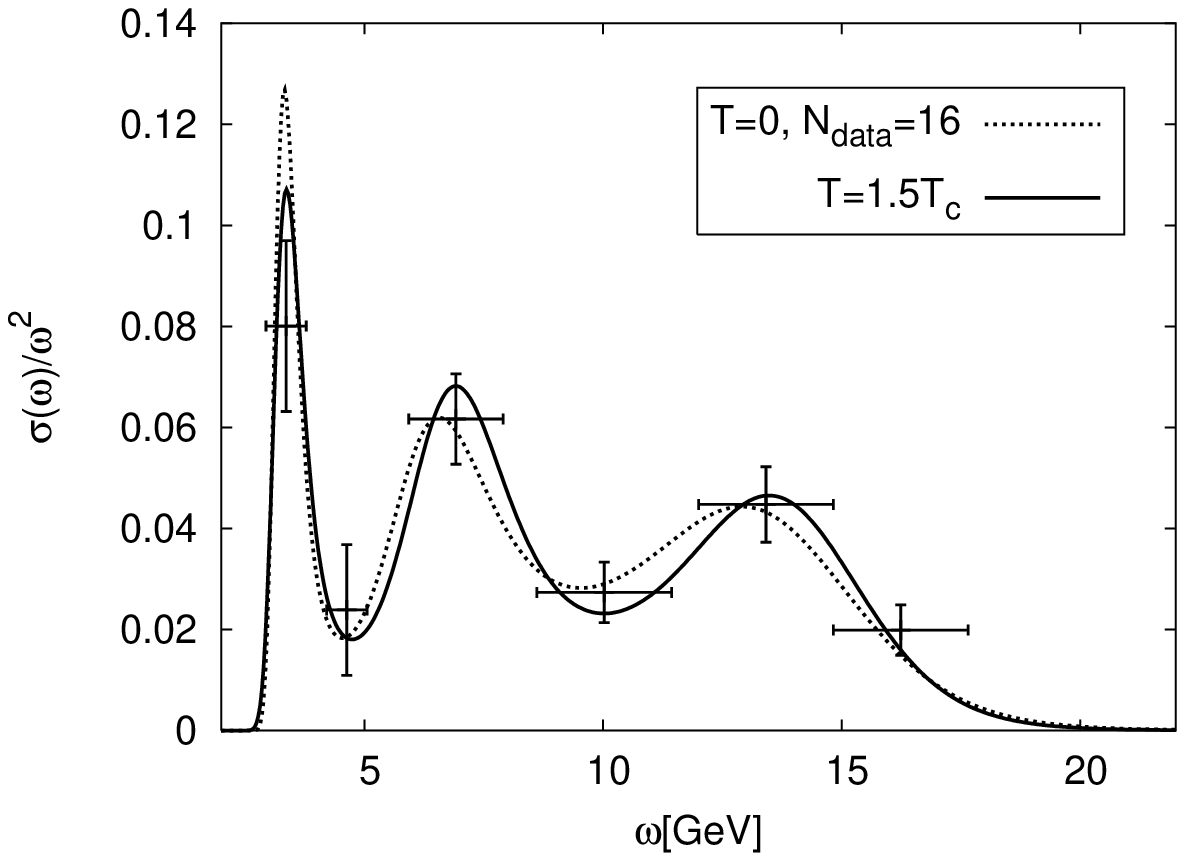,height=43mm}
\end{minipage}
\caption{ Temperature dependence of pseudoscalar charmonium correlators (left panel) and spectral functions (right panel) from the lattice \cite{Petrov:2005ej}.} 
\label{fig:lattice-c}
\end{figure}

\begin{figure}[htbp]
\hspace*{1.2cm}
\begin{minipage}[htbp]{5cm}
\epsfig{file=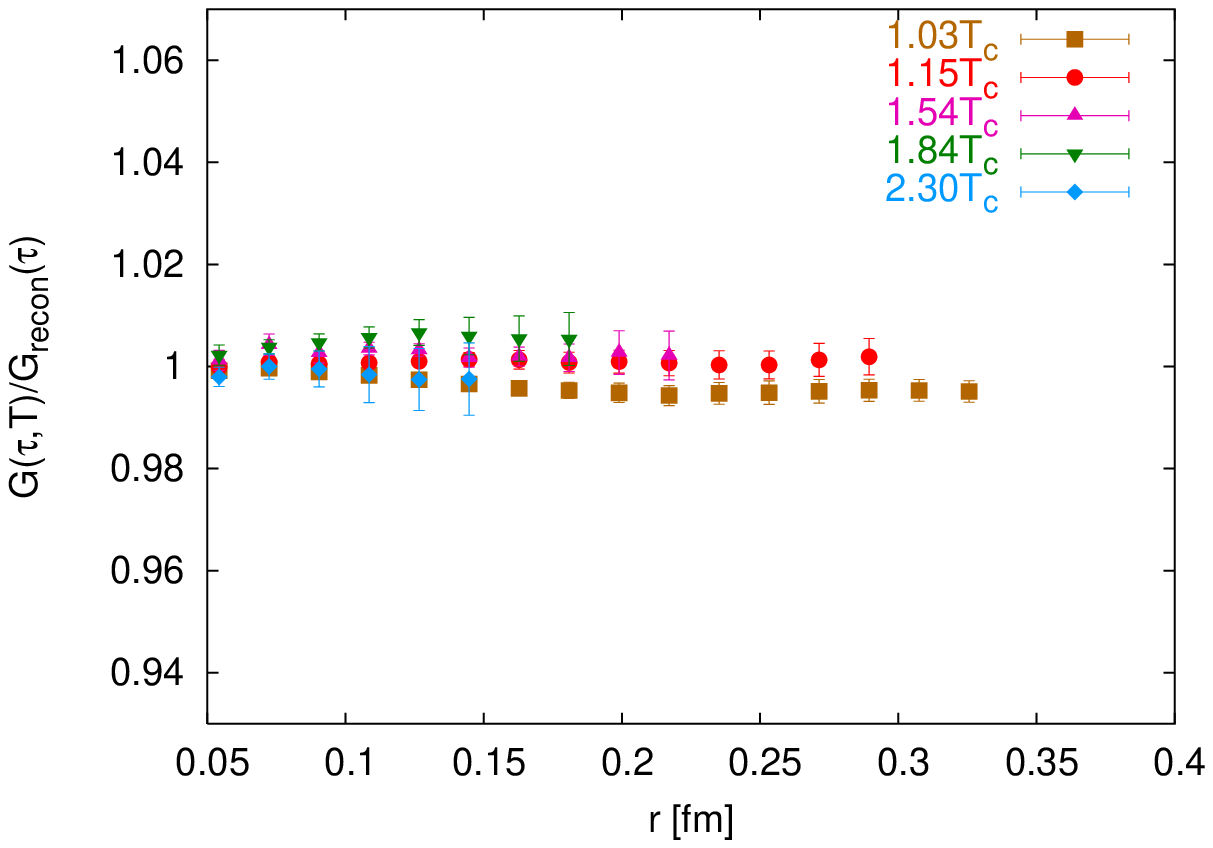,height=43mm}
\end{minipage}
\hspace*{1cm}
\begin{minipage}[htbp]{5cm}
\epsfig{file=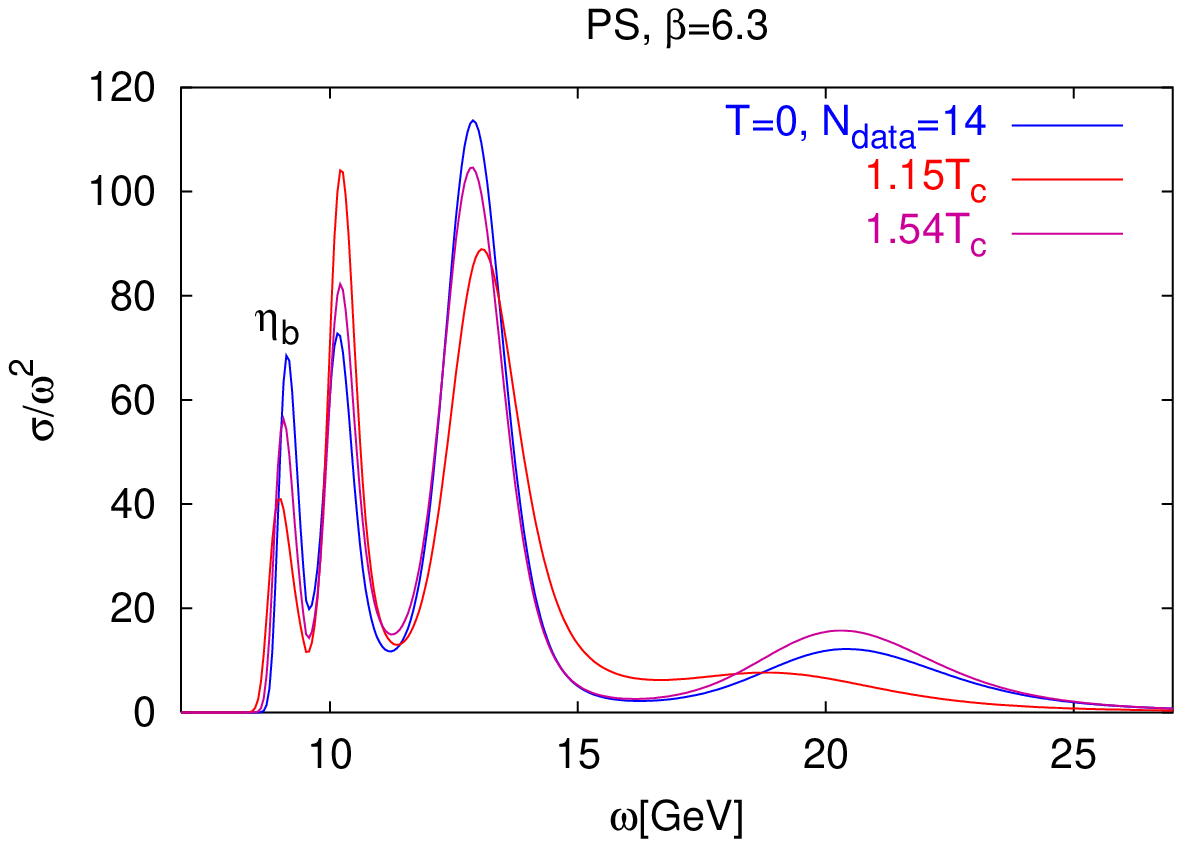,height=46mm}
\end{minipage}
\caption{Temperature dependence of pseudoscalar bottomonium correlators (left panel) spectral functions (right panel ) from the lattice \cite{Petrov:2005ej}.} 
\label{fig:lattice-b}
\end{figure}

The temperature-dependence of the meson correlators  can provide information about the fate of quarkonia states above deconfinement. The Euclidean correlation functions of meson currents $G(\tau,T)$ are reliably  calculated on the lattice
\begin{equation}
G(\tau,T)=\int_{0}^{\infty}d\omega \sigma(\omega,T)K(\omega,\tau,T)\label{corr} \, ,
\label{G}
\end{equation}
with the integration kernel $K(\tau,\omega,T) =\cosh{(\omega(\tau-1/2T))}/\sinh{(\omega/2T)}$.  The spectral functions $\sigma(\omega,T)$ are extracted from (\ref{G}) using the Maximum Entropy Method. 
The results from \cite{Petrov:2005ej} for the S-wave pseudoscalar channel are shown on Fig.~\ref{fig:lattice-c} for charmonium and on Fig.~\ref{fig:lattice-b} for bottomonium. The left panels display the  correlator normalized to the reconstructed correlator  $G_{recon}(\tau,T)=\int d\omega \sigma(\omega,T=0)K(\omega,\tau,T)$, while the right panels show the corresponding spectral functions.
One can see that there is little modification in the S-state correlators above $T_c$: the 1S charmonium state survives up to $1.5T_c$, and the 1S bottomonium shows very small change up to $2.3T_c$. These are confirmed by the spectral functions, which further suggest almost no modification of the properties of the 1S  states compared to their $T=0$ value. 

These lattice results are in contradiction with the original potential model predictions. Potential models assume that the interaction between a heavy quark and antiquark is instantaneous and is mediated by a two-body potential. The properties of a bound state are determined by solving the Schr\"odinger equation with this potential. At zero temperature the Cornell potential seems to have described quarkonia spectroscopy rather well. At finite temperature, however, the form of the potential is not known. It is even questionable whether a temperature-dependent quark-antiquark potential is adequate for the understanding of the properties of quarkonia at finite temperature. 

Recently, the potential models have been reconsidered using different temperature-dependent potentials \cite{Shuryak:2003ty,{Wong:2004zr}}. In \cite{Mocsy:2004bv}, however,  it has been shown, that even though certain screened potentials can reproduce qualitative features of the lattice spectral function, such as the survival of 1S state and the melting of 1P state, the temperature dependence of neither the meson correlators, nor the properties of the S-states is reproduced. 

\begin{figure}[htbp]
\begin{center}
\resizebox{0.40\textwidth}{!}{%
 \includegraphics{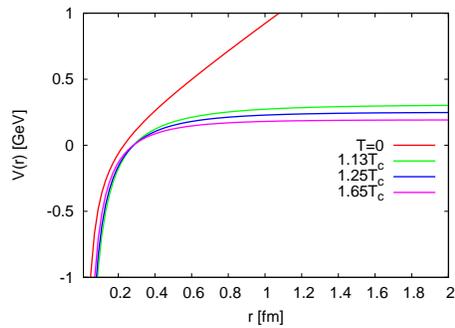}}
\caption{The Wong-potential. }
\label{fig:pot}
\end{center}
\end{figure}

Here we illustrate these findings for the potential suggested by Wong \cite{Wong:2004zr}: A screened potential constructed from the color singlet free energy and the internal energy of a heavy quark-antiquark pair in quenched lattice QCD.  This potential is shown on Fig.~\ref{fig:pot}, and its detailed derivation can be found in \cite{Wong:2004zr}. We studied the correlators and spectral functions of heavy quark bound states using, besides others, this Wong-potential. 

\begin{figure}[htbp]
\hspace*{1cm}
\begin{minipage}[htbp]{5cm}
\epsfig{file=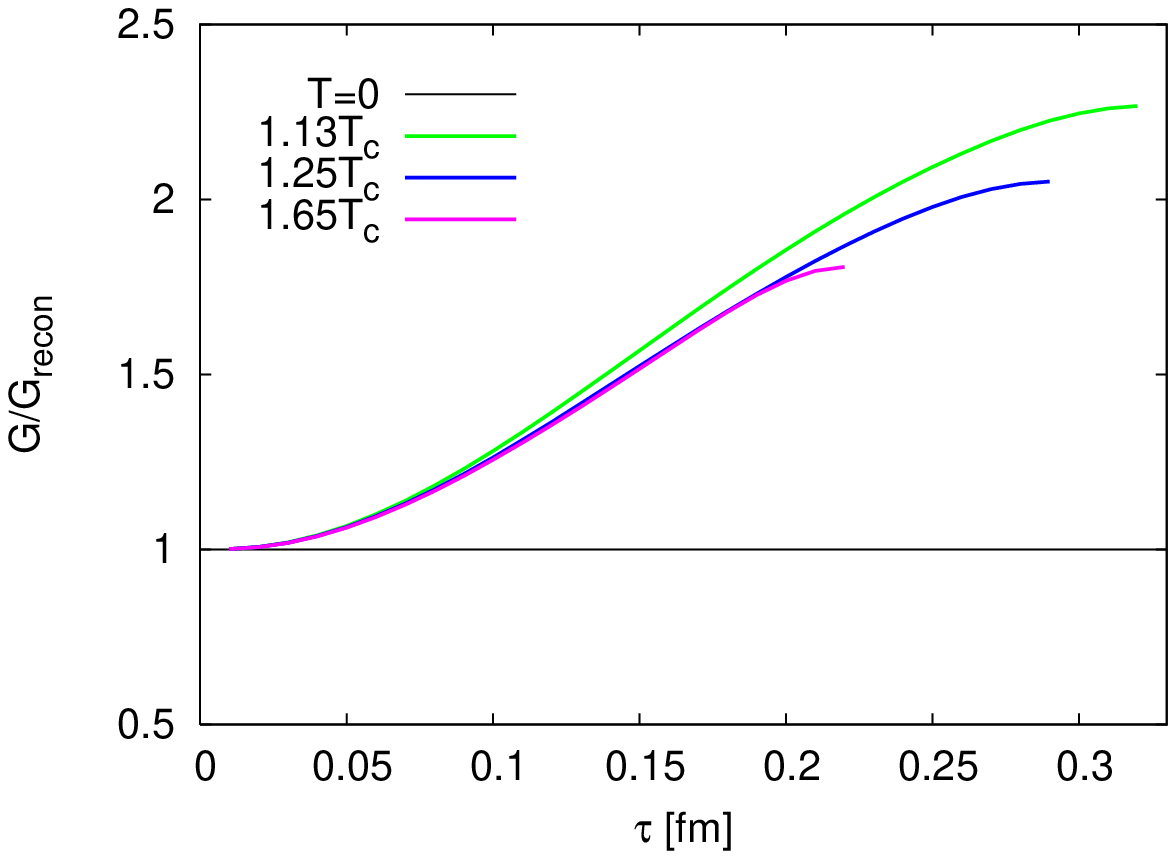,height=46mm}
\end{minipage}
\hspace*{1cm}
\begin{minipage}[htbp]{5cm}
\epsfig{file=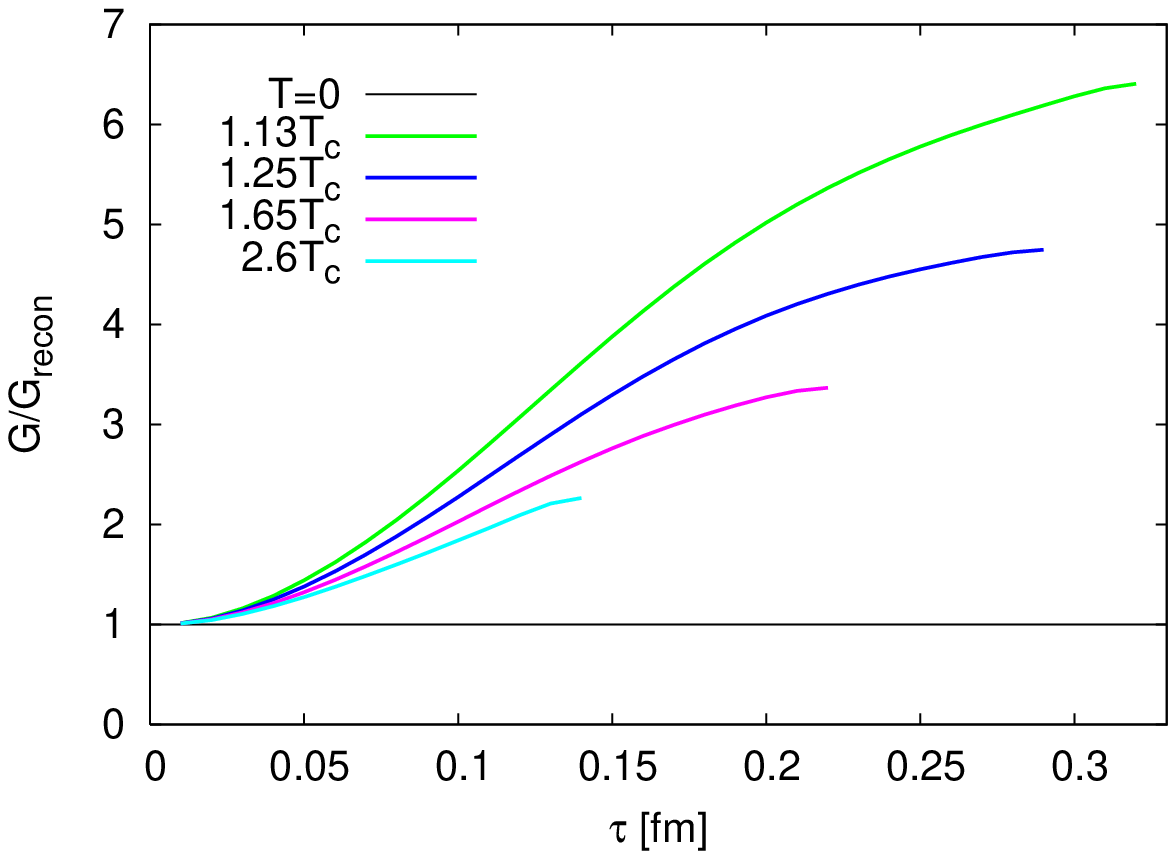,height=46mm}
\end{minipage}
\caption{S-wave charmonium (left panel) and bottomonium (right panel) correlators for different temperatures from the model spectral function.} 
\label{fig:corr}
\end{figure}

We calculated the correlator inserting in (\ref{G}) a temperature-dependent spectral function that we designed following the usual $T=0$ form, namely to have a contribution from the bound states (resonances), and a continuum above a threshold $s_0$ \cite{Mocsy:2004bv}. For the S-states this reads as
\be
\sigma(\omega,T) = \sum_i 2 M_i(T) F_i(T)^2 \delta\left(\omega^2-M_i(T)^2\right) + 3/(8\pi^2)\omega^2 \theta\left(\omega-s_0(T)\right) \, .  
\label{spft} 
\ee 
The temperature-dependent masses $M_i$ and amplitudes $F_i$ of the $c\bar{c}$ and $b\bar{b}$ states is determined by solving the Schr\"odinger equation with the Wong-potential.  In order to make direct comparison to the lattice results we calculate the ratio of the correlator (\ref{G}) to the reconstructed one. The results for the S-wave charmonium and bottomonium are presented on the left and right panels of Fig.~\ref{fig:corr}.      
One can see that neither the charmonium (compare left panels of Figs.~\ref{fig:lattice-c} and \ref{fig:corr}) nor the bottomonium (compare left panel of Fig.~\ref{fig:lattice-b} with right panel of Fig.\ref{fig:corr}) correlators reproduce the qualitative features of the correlators obtained on the lattice. The increase found in the potential model correlators indicates that the S-wave quarkonia spectral function is significantly altered compared to its $T=0$ form already at temperatures close to $T_c$. This increase suggests the possible melting of the states.

Since using other temperature-dependent screened potentials also resulted in correlators that did not reproduce the qualitative features of the lattice data \cite{Mocsy:2004bv}, we raised the question whether such potential description of the quarkonium states at finite temperature is suitable. Also, whether screening is the mechanism responsible for the dissolution of the heavy quark bound states? In a simple toy model with no temperature-dependent screening we showed that the decrease of the threshold can compensate for the melting of the higher excited states above $T_c$, and reproduce the qualitative behavior of the lattice correlators for all the states. One can easily imagine that the time-scale of screening is not short compared to the time-scale of the heavy quark motion. In this case many-body interactions with the medium become important. 

\begin{figure}[htbp]
\hspace*{1.3cm}
\begin{minipage}[htbp]{5cm}
\epsfig{file=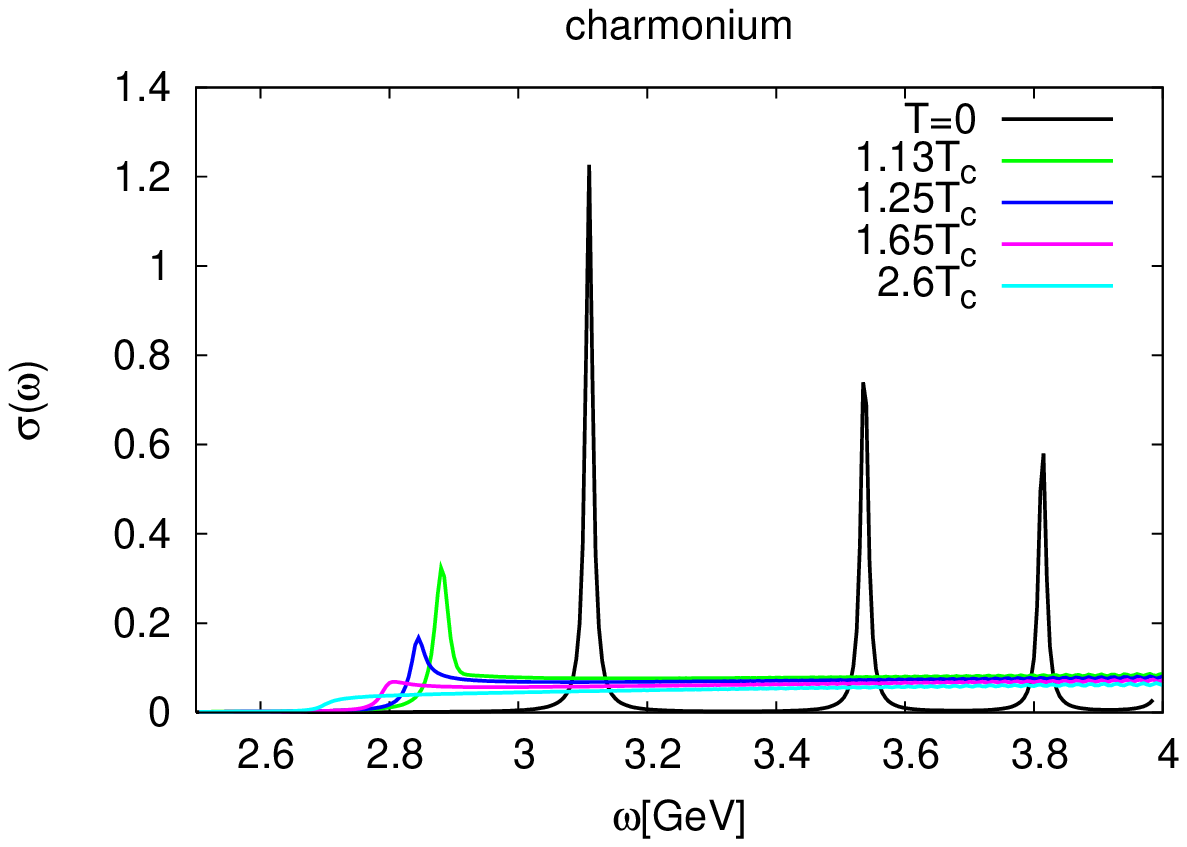,height=46mm}
\end{minipage}
\hspace*{1cm}
\begin{minipage}[htbp]{5cm}
\epsfig{file=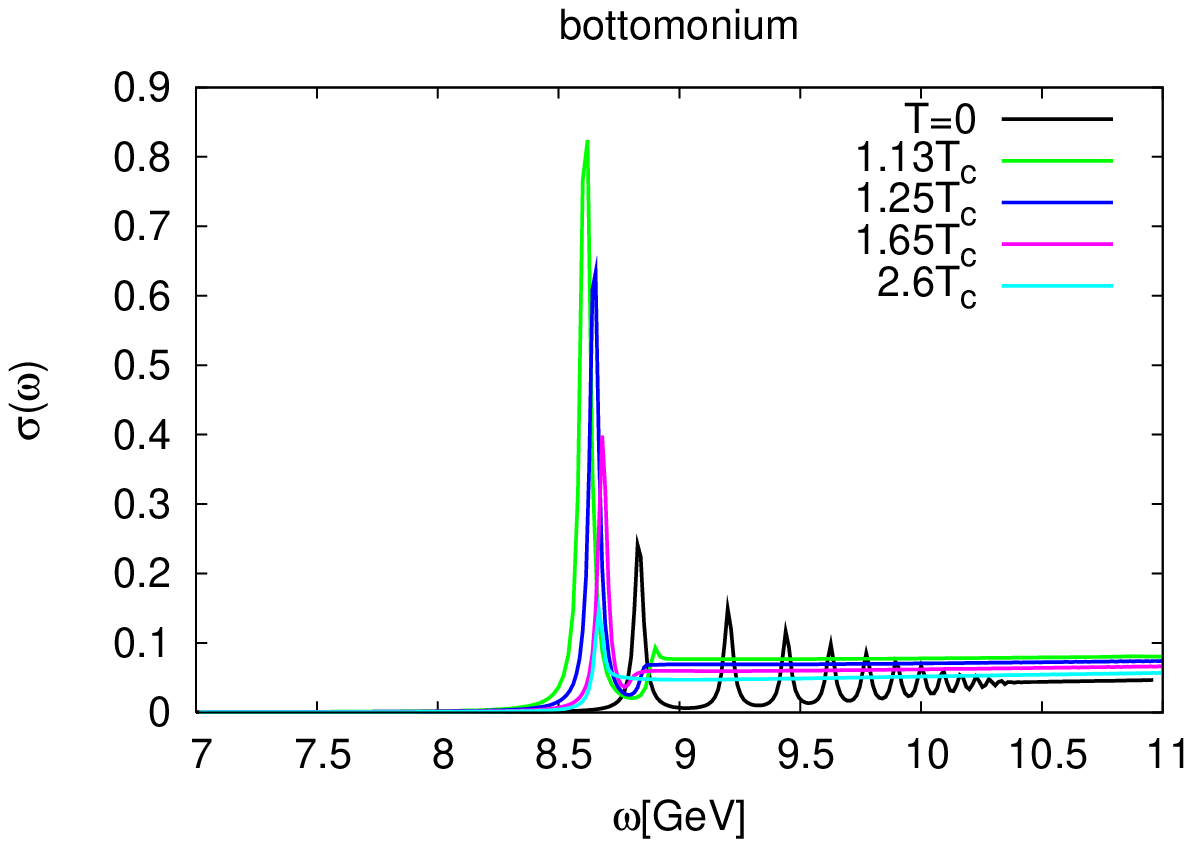,height=46mm}
\end{minipage}
\caption{S-wave charmonium (left panel) and bottomonium (right panel) spectral function for different temperatures from the nonrelativistic Green's function evaluation.} 
\label{fig:charm}
\end{figure}

The Ansatz for the spectral function given in (\ref{spft}) is justified when there is a sizable gap between the resonance and the continuum. As the temperature increases this gap becomes smaller and therefore equation (\ref{spft}) is not a very good Ansatz. To overcome this problem we performed a full non-relativistic calculation of the Green's function \cite{grennfct}, whose imaginary part provides the quarkonium spectral function.  Such an approach was used to determine properties of $t\bar{t}$ states \cite{Strassler:1990nw}, and then for binary light quark states \cite{Casalderrey-Solana:2004dc}. The results using the Wong-potential are presented in Fig.~\ref{fig:charm} for the S-wave charmonium (left panel) and bottomonium (right panel). The charmonium spectral function shows a drastic decrease of the mass and amplitude of the 1S state near $T_c$, an effect not observed on the lattice (compare to right panel of Fig.~\ref{fig:lattice-c}).  We also see a clear shift of the 1S bottomonium state, and an increase in the amplitude. These features are not seen on the lattice (compare to right panel of Fig.~\ref{fig:lattice-b}).  The approach based on non-relativistic Green's functions is not correct at large energies, where the relativistic continuum $\sigma(\omega)\sim\omega^2$ should be observed. The T-dependence of the $G/G_{recon}$ ratio, however should be correctly observed in this approach. But our calculations show no qualitative agreement of this with the lattice \cite{grennfct}.  

Based on our analysis we conclude that the potential proposed by Wong is not the correct description of the quarkonium states at high temperatures. 

\'A.M.  thanks the Organizers for the opportunity to present this work.

\end{document}